\documentclass[aps,prl,twocolumn,showpacs,amsmath,amssymb]{revtex4}
\usepackage{epsfig}
\usepackage{bm}
\newcommand{\bbb}{\!}

\begin{document}

\title{Entanglement in Anderson Nanoclusters}
\author{Peter Samuelsson}
\affiliation{Solid State Theory,  Lund University, S\"olvegatan 14 A, 223 62 Lund, Sweden}
\author{Claudio Verdozzi}
\affiliation{Solid State Theory,  Lund University, S\"olvegatan 14 A, 223 62 Lund, Sweden}
\date{\today}

\begin{abstract}
We investigate the two-particle spin entanglement in magnetic
nanoclusters described by the periodic Anderson model. An entanglement
phase diagram is obtained, providing a novel perspective on a central
property of magnetic nanoclusters, namely the temperature dependent
competition between local Kondo screening and nonlocal
Ruderman-Kittel-Kasuya-Yoshida spin ordering.  We find that
multiparticle entangled states are present for finite magnetic field
as well as in the mixed valence regime and away from half filling. Our
results emphasize the role of charge fluctuations.
\end{abstract}

\maketitle

In the last decade, solid state systems have emerged as a promising
stage for quantum information processing, due to the prospect of
scalability and integrability with conventional electronics. A
successful realization of solid state quantum information processing
requires a detailed control of the quantum mechanical properties of
the system. In this respect, a key property is entanglement, or
quantum mechanical correlations, between the individual quantum bits;
entanglement plays the role of basic resource for a large number of
quantum information schemes \cite{NielsenChuang}. In nanoscale solid
state systems, internal degrees of freedom of individual electrons,
like spin or orbital states, are natural candidates for quantum bits.
This offers considerable scope for studies of electronic entanglement
in nanosystems.

For spin entanglement, of particular interest are nanoclusters with
magnetic impurities, realized experimentally e.g. by magnetic atoms on
a surface \cite{Manoharan}, on a nanotube \cite{Odom} or by coupled
quantum dots \cite{Craig}. Such nanoclusters display intriguing spin
properties as the Kondo effect and antiferromagnetism, similar to what
occurs in extended systems with dense magnetic impurities.  For
extended systems a central feature, described by the Doniach phase
diagram \cite{Doniach}, is the competition between formation of local
Kondo spin singlets and a nonlocal Ruderman-Kittel-Kasuya-Yoshida
(RKKY) spin ordering. A model which captures such behavior is the
periodic Anderson model, PAM, a lattice of localized levels (with
strong local repulsion) which hybridize with a conduction band. The
PAM has been investigated intensively during the last decades
\cite{Sigrist} in connection with heavy fermion physics, non-fermi
liquid behavior, etc \cite{Kondoref}.

To date, no studies are however available for entanglement in the PAM,
only the related single \cite{Kim} and two \cite{Cho} impurity
Anderson models and the simplified Kondo necklace model \cite{Saguia}
have been considered. There are also investigations of entanglement in
various spin cluster models \cite{spinmod,Arnesen}. None of these
models do however correctly capture the interplay of spin correlations
and charge dynamics, characteristic of the PAM \cite{Sigrist}. Only
very recently the interest turned to systems where charge dynamics is
important \cite{Gu,Johannesson}. These investigations however only
concerned the Hubbard model, focusing on entanglement in extended
systems, at quantum phase transitions \cite{Fazio}.  An investigation
of the entanglement in nanoclusters described by the PAM is thus
highly desirable.
\begin{figure}[b]
\includegraphics*[width=.37\textwidth]{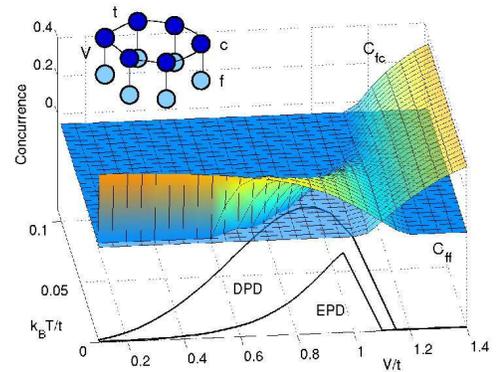}
\caption{ The concurrences $C_{\text{ff}}^{(1)}$ and
$C_{\text{fc}}^{(0)}$ as a function of $k_{B}T/t$ and $V/t$ for $L=6$,
$U=-2E_{\text{f}}=6t$, $B=0$ and half filling. Projected at the bottom is the
Doniach, DPD,  and the corresponding entanglement phase diagram, EPD 
(see text). The $L=6$ cluster is also shown.}
\label{fig1}
\end{figure}

In this work we perform such an investigation. Using exact
diagonalization methods, ED, we study two-particle spin entanglement
in the ground state and at finite temperatures. Both the Kondo- (well
localized moments) and the mixed valence- (charge fluctuations)
regimes are investigated. The effect of magnetic field and different
electron fillings are considered. We find that: (i) The entanglement
is governed by the hybridization, i.e. by the coupling between the
conduction (c-) electrons and the local moments due to the f-electrons
(see Fig. \ref{fig1}). At small hybridization the entanglement is
predominantly nonlocal, between f-electrons, while at large
hybridization it is local, between f- and c-electrons. The local
fc-entanglement survives at much higher temperatures than does the
nonlocal ff-entanglement. (ii) These properties can be represented by
an entanglement phase diagram, providing a view on the Kondo-RKKY
competition complementary to the Doniach phase diagram. (iii) In the
mixed valence regime, at finite magnetic field and away from half
filling we find nontrivial multiparticle entangled states. (iv) Charge
dynamics plays an essential role in obtaining the results (i) to
(iii).  Our results illustrate the richness of entanglement properties
of the PAM and invite to further investigations of additional regions
in parameter space, disorder and geometry effects, etc.

{\it Model and theory} We study ring shaped clusters with $L=2$ to $6$
sites and two orbitals per site, denoted c and f (see Fig. \ref{fig1}
for $L=6$). Each cluster has $N_{e}$ electrons, with $0\le N_{e}\le
4L$.  The PAM cluster Hamiltonian is
\begin{eqnarray}
H&=&-t\sum_{\langle i,j \rangle\sigma}(c_{i\sigma}^{\dagger}c_{j\sigma}+h.c.)+U\sum_{i}n_{i\uparrow}^fn_{i\downarrow}^f+E_f\sum_{i\sigma}n_{i\sigma}^f\nonumber\\
&+&V\sum_{i\sigma}(f_{i\sigma}^{\dagger}c_{i\sigma}+h.c.)+g\mu_BB
\sum_{i\alpha}S_{i\alpha}^{z}
\label{ham}
\end{eqnarray}
with $n_{i\sigma}^f=f_{i\sigma}^{\dagger}f_{i\sigma}$ and
$\sigma=\uparrow,\downarrow$ , $\alpha=c,f$. Here $\langle i,j\rangle$
denotes nearest neighbor, n.n., sites and the spin operator $\hat
S_{ic}=(S_{ic}^x,S_{ic}^y,S_{ic}^z)=(1/2)\sum_{\sigma,\sigma'}\hat
\tau_{\sigma\sigma'}c_{i\sigma}^{\dagger}c_{i\sigma'}$ for c
(similarly for f), with $\hat \tau=(\tau_x,\tau_y,\tau_z)$ a vector of
Pauli matrices. The n.n. hopping between c-orbitals is $t$ ($t>0$),
whilst $V$ is the hybridization term between c and f-orbitals at the
same site: both $t$ and $V$ can be taken real.  The f-orbital onsite
interaction strength is $U$, $E_f$ is the energy of the f-orbitals and
$B$ is the magnitude of the uniform magnetic field.

We focus on the reduced two particle spin density matrix $\rho\equiv
\rho_{\alpha\beta}^{(|i-j|)}$, for orbitals $\alpha,\beta=f,c$ at
sites $i$ and $j$. Using the $\{|\!\!
\uparrow\uparrow\rangle,|\!\!\uparrow\downarrow\rangle,|\!\!
\downarrow\uparrow\rangle,|\!\!\downarrow\downarrow\rangle\}$ basis,
for e.g. $\alpha,\beta=f$ the density matrix elements are $\langle
f_{i\sigma}^{\dagger}f_{j\sigma'}^{\dagger}f_{j\sigma''}f_{i\sigma'''}\rangle$
where $\langle .. \rangle$ denotes exact, many-body, thermal
equilibrium averages. Due to translational invariance $\rho$ depends
only on $|i-j|$. Since $[H,\sum_{i\alpha} S_{i\alpha}^{z}]=0$, each
eigenstate has a well defined number of spin up and down electrons and
only spin-conserving density matrix elements are nonzero.
Consequently, $\rho$ has in general five independent parameters:
\begin{equation}
\rho=\left(\begin{array}{cccc} a &0&0&0 \\ 0 & b & c & 0 \\ 0& c &
\tilde{b} &0 \\ 0 &0&0&\tilde {a} \end{array}\right).
\label{densmat}
\end{equation}
Note that typically $\mbox{tr}(\rho)\neq 1$, since the number of
electrons at each orbital can vary from zero to two. When $B=0$ spin
symmetry requires\cite{Nozieres} the state to be characterized by only
two free parameters, $c$ and $a$, with $a=\tilde a$ and $b=\tilde
b=a-c$: the state is a Werner state \cite{Werner}.

The entanglement is conveniently quantified via the concurrence
\cite{Wooters} $ C(\rho)\equiv C^{(|i-j|)}_{\alpha\beta}(\rho)=
\mbox{max}\{0,\sqrt{\lambda_1}-\sqrt{\lambda_2}-\sqrt{\lambda_3}-\sqrt{\lambda_3}\}$
where the $\lambda_i$s are the real and positive eigenvalues, in
decreasing order, of the matrix $\rho\tilde \rho$, where $\tilde
\rho=(\tau_y\otimes \tau_y)\rho^*(\tau_y\otimes \tau_y)$.  This gives
\cite{conccom} for $\rho$ in Eq. (\ref{densmat})
\begin{equation}
C=2~\mbox{max}\{c-\sqrt{a\tilde a},0\} , \hspace{0.5 cm} 0\leq C \leq 1
\label{conceq}
\end{equation}
We point out that in contrast to the single site Fock space, or mode
entanglement considered e.g. in \cite{Gu,Johannesson}, here the
entanglement between two physical particles \cite{Wiseman} is
considered. The concurrence $C$ in Eq. (\ref{conceq}) is obtained from
the reduced density matrix $\rho$ which by definition determines any
two-particle observable, as e.g. correlation functions. Thus,
$C(\rho)$ is a natural measure for the experimentally accessible two
particle entanglement.

{\it Kondo-RKKY competition} A central property of the PAM in the
local moment (Kondo) regime is the competition between Kondo and RKKY
correlations for the f-electrons. For macroscopic systems, such
competition was described qualitatively by Doniach \cite{Doniach}: for
low temperature $T$ and weak hybridization, $V/t<1$, the localized
f-electron spins are RKKY-ordered. Increasing $V/t$ there is a cross
over to local Kondo screening. Such competition occurs also in
nanoclusters, where it is controlled both by $V/t$ and the conduction
level spacing \cite{verdozzidis}.  A description in terms of a cluster
Doniach diagram is possible \cite{Verdozzi}: A transition criterion
was established by comparing spin correlators.  Recently, similar
results have been presented for the 2D Kondo lattice model
\cite{KLM2D}.

To investigate the entanglement in the Kondo-RKKY competition, we
consider the symmetric case $U=-2E_f$ (the ground state
$|\Psi_{\text{g}}\rangle$ is a singlet \cite{Sigrist}), with well
localized f-electrons at half filling ($N_{e}=2L$), $B=0$ and
$T=0$. Under these conditions the PAM can be mapped
\cite{SchriefferWolf} onto the Kondo lattice model (KLM)
\cite{Sigrist}, characterized solely by $\tilde{J}=8V^2/(Ut)$. In the
KLM both the ff and fc density matrices are normalized and thus
parametrized by a single parameter $a$. The concurrence is given by
$C=\mbox{max}\{1-6a,0\}$ and can be directly related to the spin
correlator $K\equiv K_{\alpha\beta}^{(|i-j|)}=\langle S_{i
\alpha}^zS_{j \beta}^z\rangle$ as $C=\mbox{max}\{-1/2-6K,0\}$, where
$K=a-1/4$.

\begin{figure}[tbp]
\centerline{\psfig{figure=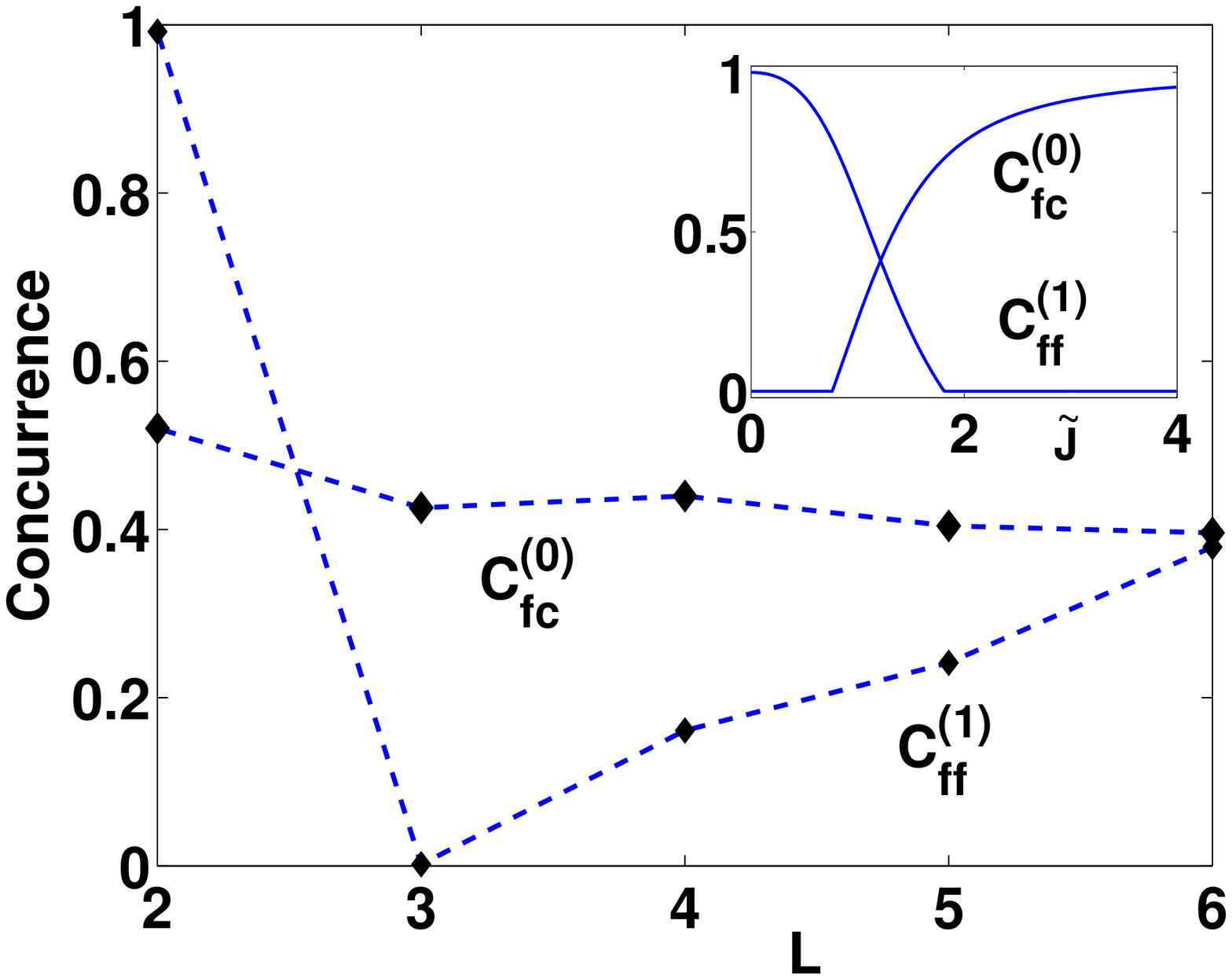,width=3.8cm}\psfig{figure=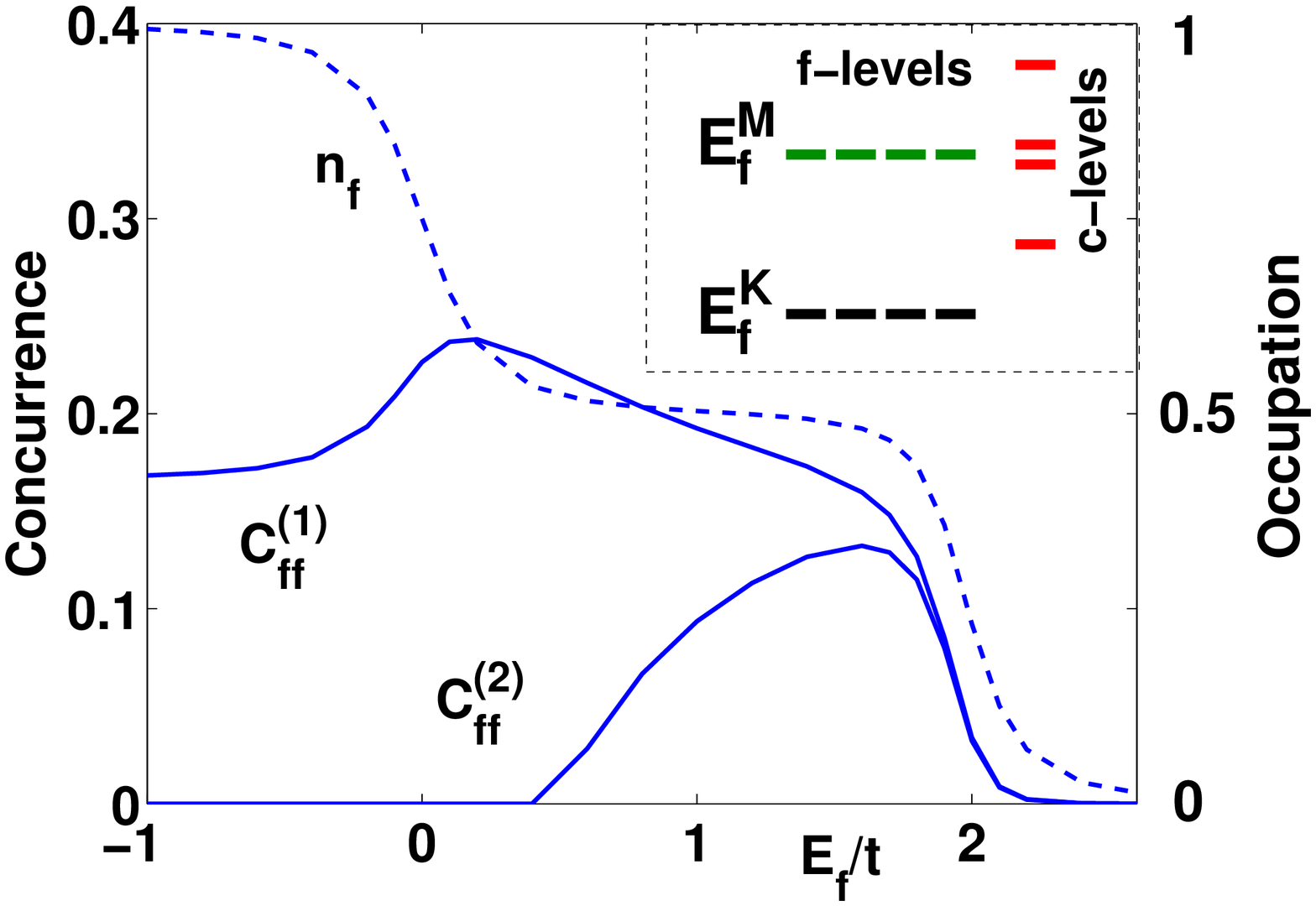,width=4.3cm}}
\caption{ Left: Concurrence $C_{\text{ff}}^{(1)}$ (for $V=0.2t$) and
$C_{\text{fc}}^{(0)}$ (for $V=1.5t$) as a function of system size
$L$. Inset: $C_{\text{ff}}^{(1)}$ and $C_{\text{fc}}^{(0)}$ as a
function of $\tilde J$ for the $L=2$ KLM. Right: $C_{\text{ff}}^{(1)}$
and $C_{\text{ff}}^{(2)}$ and the total f-level occupation $n_f$ as a
function of $E_{\text{f}}/t$ in the mixed valence regime,
$V=0.1t,U=6t$ and $L=4$. The energy level scheme for $V \ll t$ is
shown as inset: $E_{\text{f}}^M$ ($E_{\text{f}}^K$) denotes the
f-level in the mixed valence (Kondo) regime.}
\label{fig2}
\vspace{-0.5cm}
\end{figure}

Preliminary insight in the Kondo-RKKY competition is gained from the
KLM in the simple case $L=2$, where $|\Psi_{\text{g}}\rangle$ is
obtained analytically. For the two limiting cases,
\begin{equation}
\footnotesize{
|\Psi_{\text{g}}\rangle=
\left\{
\begin{array}{lr}
\frac{1}{\sqrt{8}}(f_{1\uparrow}^{\dagger}f_{2\downarrow}^{\dagger}-f_{1\downarrow}^{\dagger}f_{2\uparrow}^{\dagger})(c_{1\uparrow}^{\dagger}+c_{2\uparrow}^{\dagger})(c_{1\downarrow}^{\dagger}+c_{2\downarrow}^{\dagger})|0\rangle
&\tilde{J}\rightarrow 0\\
(1/2)(f_{1\uparrow}^{\dagger}c_{1\downarrow}^{\dagger}-f_{1\downarrow}^{\dagger}c_{1\uparrow}^{\dagger})(f_{2\uparrow}^{\dagger}c_{2\downarrow}^{\dagger}-f_{2\downarrow}^{\dagger}c_{2\uparrow}^{\dagger})|0\rangle
&\tilde{J}\rightarrow \infty
\end{array}
\right.}\normalsize{}
\end{equation}
Thus, on increasing $\tilde{J}$ from $0$ (small hybridization), the
system evolves from a maximally entangled, singlet ff-state to a
product state of cf-singlets (see Fig. \ref{fig2}). At
$\tilde{J}\approx 1.22$ we have
$C_{\text{ff}}^{(1)}=C_{\text{fc}}^{(0)}$. Qualitatively this is
similar to what occurs in the two-impurity Kondo \cite{Cho} and Kondo
necklace \cite{Saguia} models.

For $L \geq 3$ (using numerical ED with $U=-2E_f=6t$), the physical
picture for $V<t$ is very different from the $L=2$ case: The
properties of $|\Psi_{\text{g}}\rangle$ depend strongly on size
effects such as even-odd $L$ parity and the density and configuration
of the $c$-electron levels. This is clearly illustrated by the drop of
$C_{\text{ff}}^{(1)}$ from $1$ for $L=2$ to $0$ for $L=3$, shown in
Fig. \ref{fig2}. Interestingly, between $L=3$ and $L=6$,
$C_{\text{ff}}^{(1)}$ increases monotonically with $L$ from $0$ to
$0.4$. This behavior follows from an increasing weight in
$|\Psi_{\text{g}}\rangle$ of a superposition of a component with
anti-ferromagnetic like order (not perfect for odd $L$) $\sim
|\uparrow\downarrow\uparrow\downarrow....\rangle$ and the same
component with two neighboring spins flipped.

We also note that, for $L=4$ to $6$, $C^{(2)}_{\text{ff}}=0$ while for
$L=6$ second next n.n. concurrence is again finite,
$C_{\text{ff}}^{(3)}\approx 0.15$. The nonlocal concurrences
$C_{\text{fc}}^{(j)}, C_{\text{cc}}^{(j)} $ with $j\geq 1$ are
typically small or zero and will not be discussed further.
It is important to note that $|\Psi_{\text{g}}\rangle$ always contains
some doubly occupied $c$-orbitals, making the result qualitatively
different from the Kondo necklace model \cite{Saguia}. Also, an
effective RKKY spin Hamiltonian \cite{Sigrist} $\sum_{i\neq
j}J_{ij}\hat S_{if}\cdot \hat S_{jf}$ gives a very different
$|\Psi_{\text{g}}\rangle$, and hence concurrence for most $L$. This
provides strong evidence that the c-electron dynamics is essential for
the obtained results. Both ff- and fc-entanglement decrease
monotonically on increasing the temperature $T$, since unentangled
excited states get progressively populated. At $k_{B}T \sim 0.1 t$
(the typical energy scale of the low-lying spin excitations at $V <
t$) ff-entanglement is suppressed, whilst fc-entanglement persists to
much higher $T$.

A global picture of the entanglement properties of the system is
provided by an entanglement phase diagram, shown for $L=6$ in
Fig. \ref{fig1}. The phase boundary shows the crossover temperature
for which $C_{\text{ff}}=C_{\text{fc}}$. Importantly, for $V<t$ the
entanglement and the Doniach phase diagram (also in Fig. \ref{fig1})
have distinctly different crossover temperatures. Increasing $T$ from
$0$ the system thus passes three different phases: i) finite
ff-entanglement and dominating ff-spin correlators
$|K_{\text{ff}}^{(1)}|>|K_{\text{fc}}^{(0)}|$, ii) zero entanglement
and $|K_{\text{ff}}^{(1)}|>|K_{\text{fc}}^{(0)}|$ and iii) zero
entanglement and dominating fc-correlations
$|K_{\text{fc}}^{(0)}|>|K_{\text{ff}}^{(1)}|$.  Moreover, there is a
sharp cusp in the entanglement diagram due to the nonanalyticity of
the concurrence at $C=0$. This illustrates the importance of the
entanglement phase diagram \cite{tcom} as a new tool for analyzing the
Kondo-RKKY competition.

{\it Mixed valence regime}.  The effect of f-electron charge
fluctuations on the entanglement manifests quite clearly when moving
away from the Kondo regime. In the Kondo limit the f-level energies
are well below the conduction band, $|E_{\text{f}}| \gg t,V$ and the
occupation $n_f=\langle n^f_{\uparrow}\rangle+\langle
n^f_{\downarrow}\rangle$ is essentially one, due to the large value of
$U$ ($=6t$) considered.  However, when $E_{\text{f}}$ enters the
conduction band, the regime of mixed valence, $n_f$ starts to
decrease.  We studied the mixed valence entanglement for
$-1<E_{\text{f}}/t<3$, i.e. with $ n_{f} \le 1$ (deep below the
conduction band, $E_{f}+U<-2t$, the f-levels are doubly occupied and
inert).

We find that only for $V<t$ there is additional entanglment away from
the Kondo regime. In Fig. \ref{fig2}, we plot $C_{\text{ff}}$ as a
function of $E_{\text{f}}/t$ for $L=4$. When $n_{f}$ starts to drop,
both n.n. and next n.n. entanglement increase, with a maximum at
$E_{\text{f}} \approx 0.2t$ and $1.7t$, respectively. This can be
understood from the manybody wavefunction around $E_{\text{f}}\sim t$,
where the f-level occupation is roughly $1/2$, i.e. on average two
electrons occupy the f-levels.  An analysis of the exact wavefunction
shows that to a good approximation $|\Psi_{g}\rangle$ can be written
as
$|\Psi_{g}\rangle=|\Psi_{\text{f}}\rangle\otimes|\Psi_{\text{c}}\rangle$,
with
\begin{equation}
|\Psi_{\text{f}}\rangle=\frac{1}{\sqrt{12}}\sum_{i\neq j}\left(f_{i\uparrow}^{\dagger}f_{j\downarrow}^{\dagger}-f_{i\downarrow}^{\dagger}f_{j\uparrow}^{\dagger}\right)|0\rangle,\label{cinque}
\end{equation}
a linear combination of f-electron singlets. The
$|\Psi_{\text{f}}\rangle$ in Eq. (\ref{cinque}) gives a concurrence
$C_{\text{ff}}^{(1)}=C_{\text{ff}}^{(2)}=1/6$, in reasonable agreement
with the ED results in Fig. \ref{fig2}. The value $1/6$ comes entirely
from the low occupation of the f-levels, $\mbox{tr}(\rho)=1/6$, since
the singlet itself is maximally entangled. As in the Kondo regime, the
entanglement is suppressed at $k_{B}T\sim 0.1t$. The same result is
found for all $L$, with the $E_{\text{f}}/t$ interval of additional
entanglement given by the density of states in the conduction band.

{\it Finite B-field} Applying a magnetic field $B$, a naive
expectation would be that aligning more and more spins along the
direction of the field would progressively suppress the entanglement
towards zero. It was however found in spin clusters \cite{Arnesen}
that at strong magnetic fields, with all spins but one flipped, the
state is $\sim |\!\!\!\downarrow\uparrow\uparrow\uparrow...\rangle +
|\!\!\uparrow\downarrow\uparrow\uparrow...\rangle+..$, a strongly
multiparticle entangled W-state \cite{Wstate}. This motivates an
investigation the effect of $B$ on the entanglement in the PAM.

\begin{figure}[tpb]
\centerline{\psfig{figure=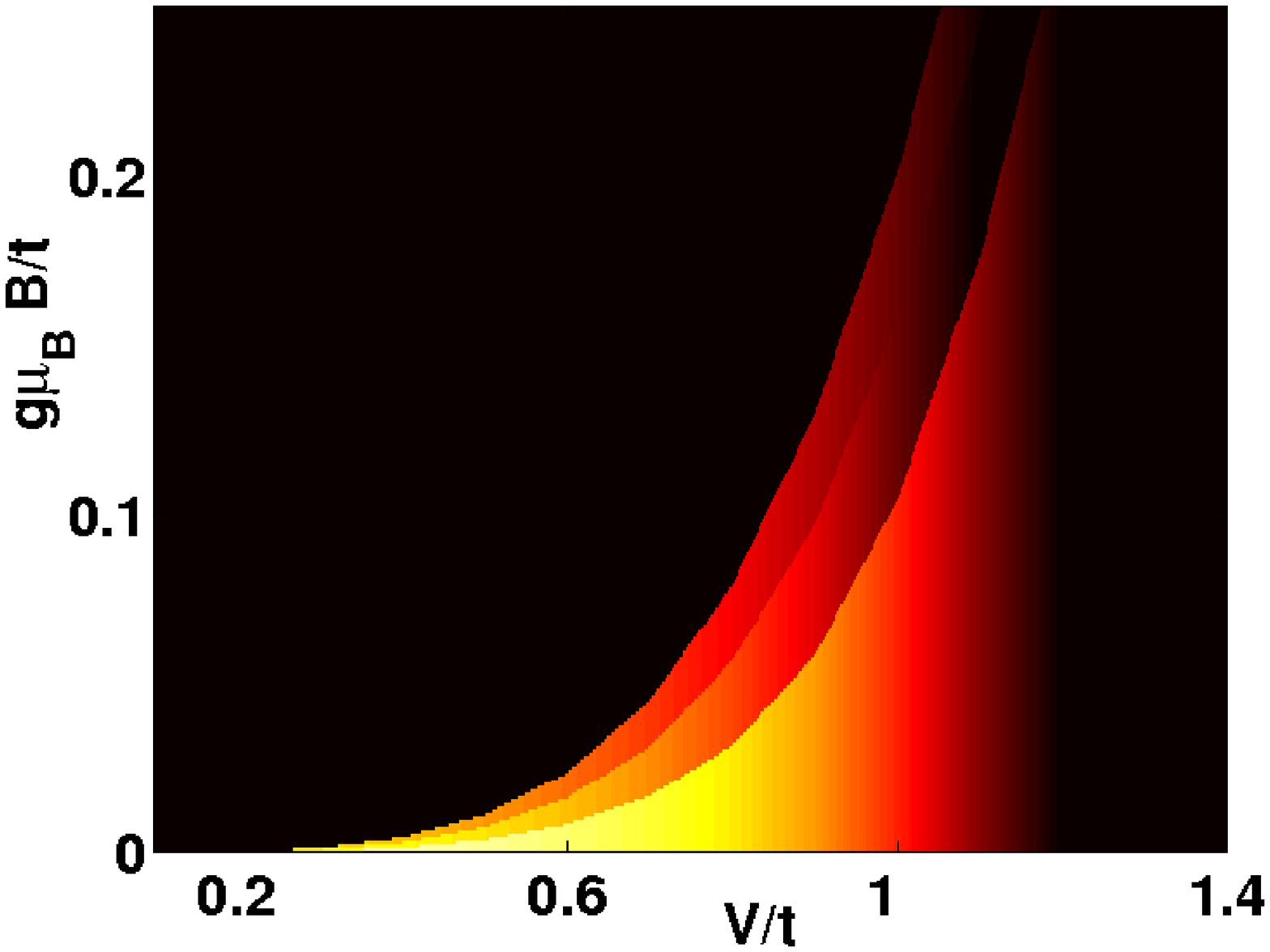,width=4.25cm}\psfig{figure=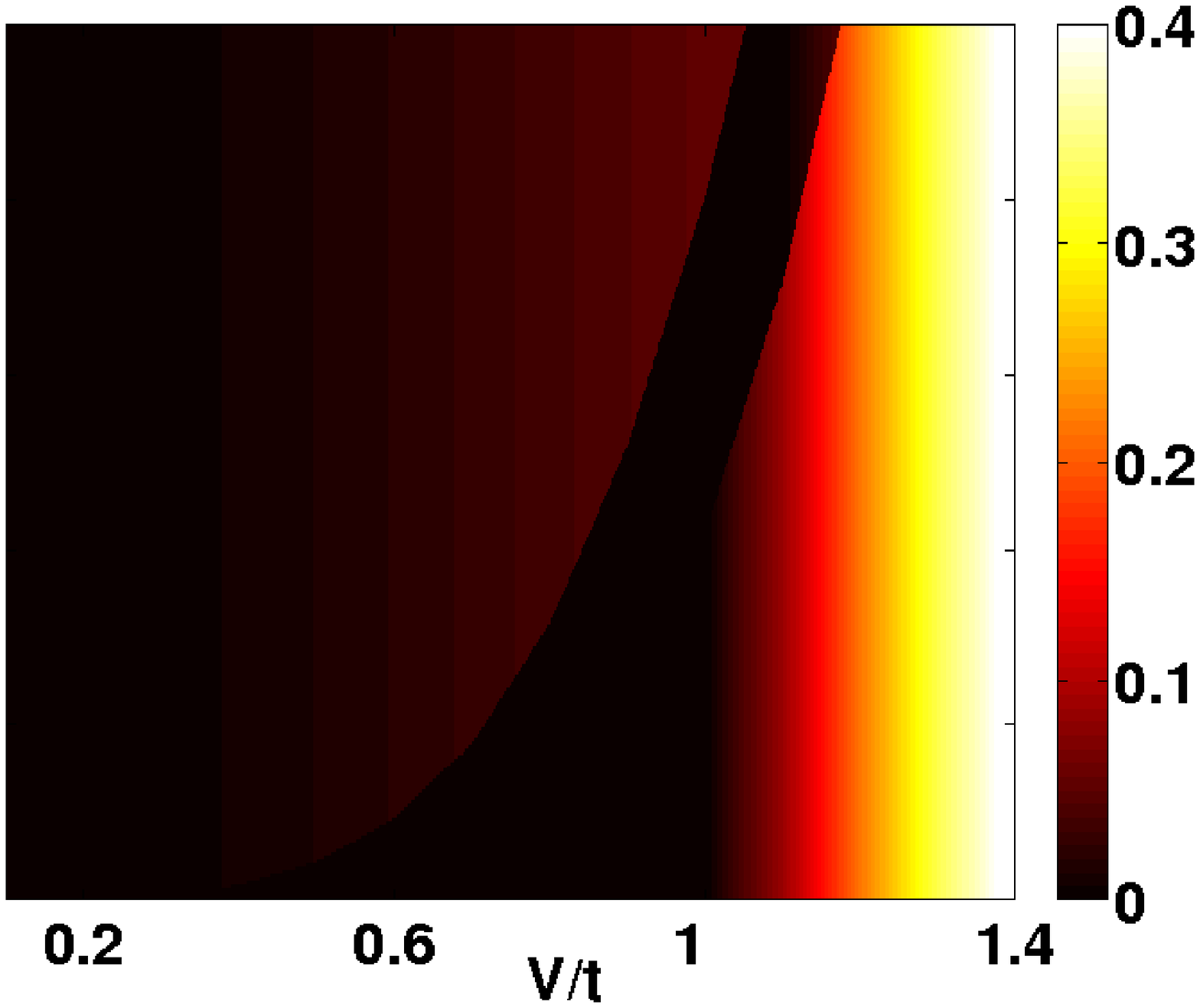,width=3.8cm}}
\centerline{\psfig{figure=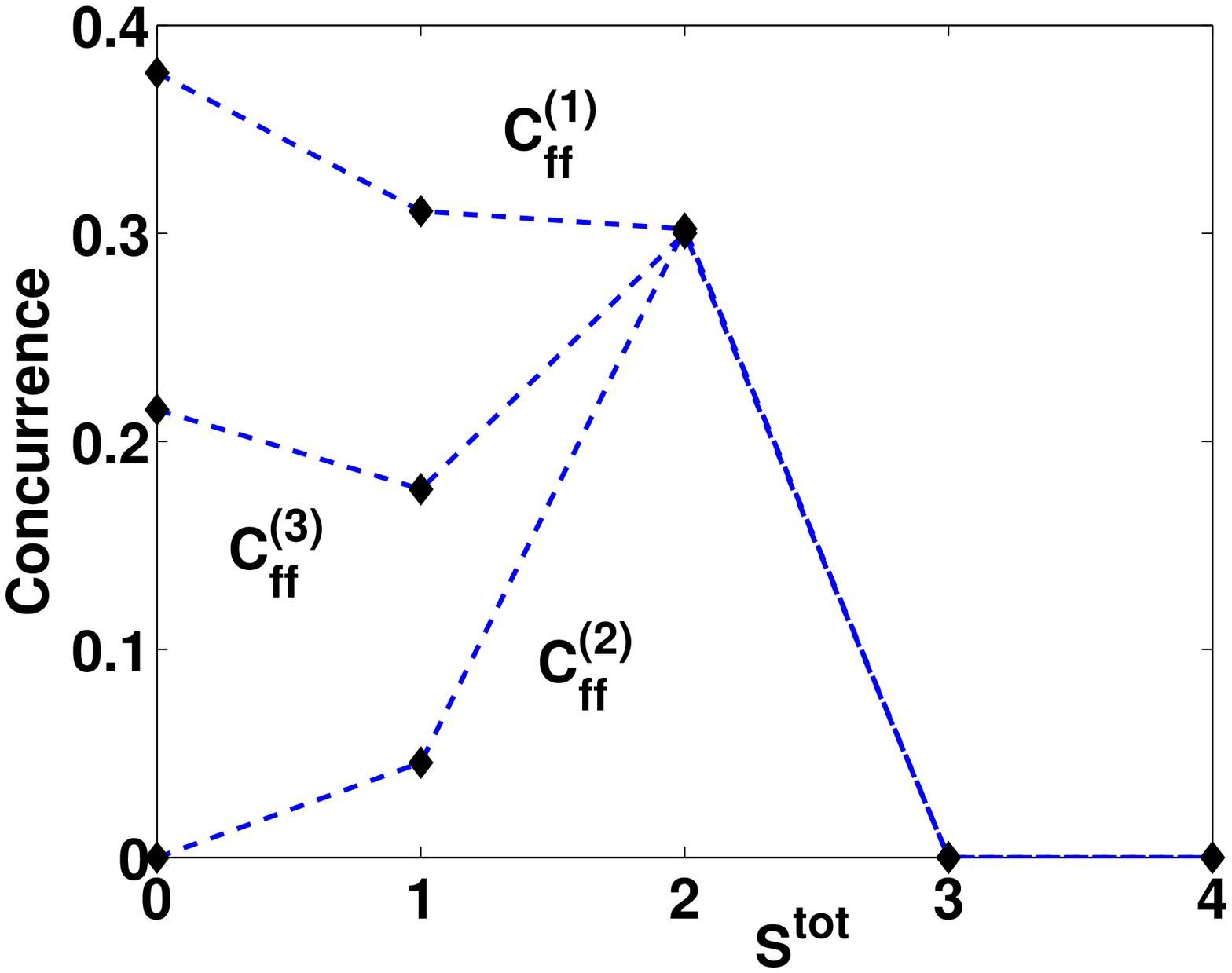,width=4cm}\psfig{figure=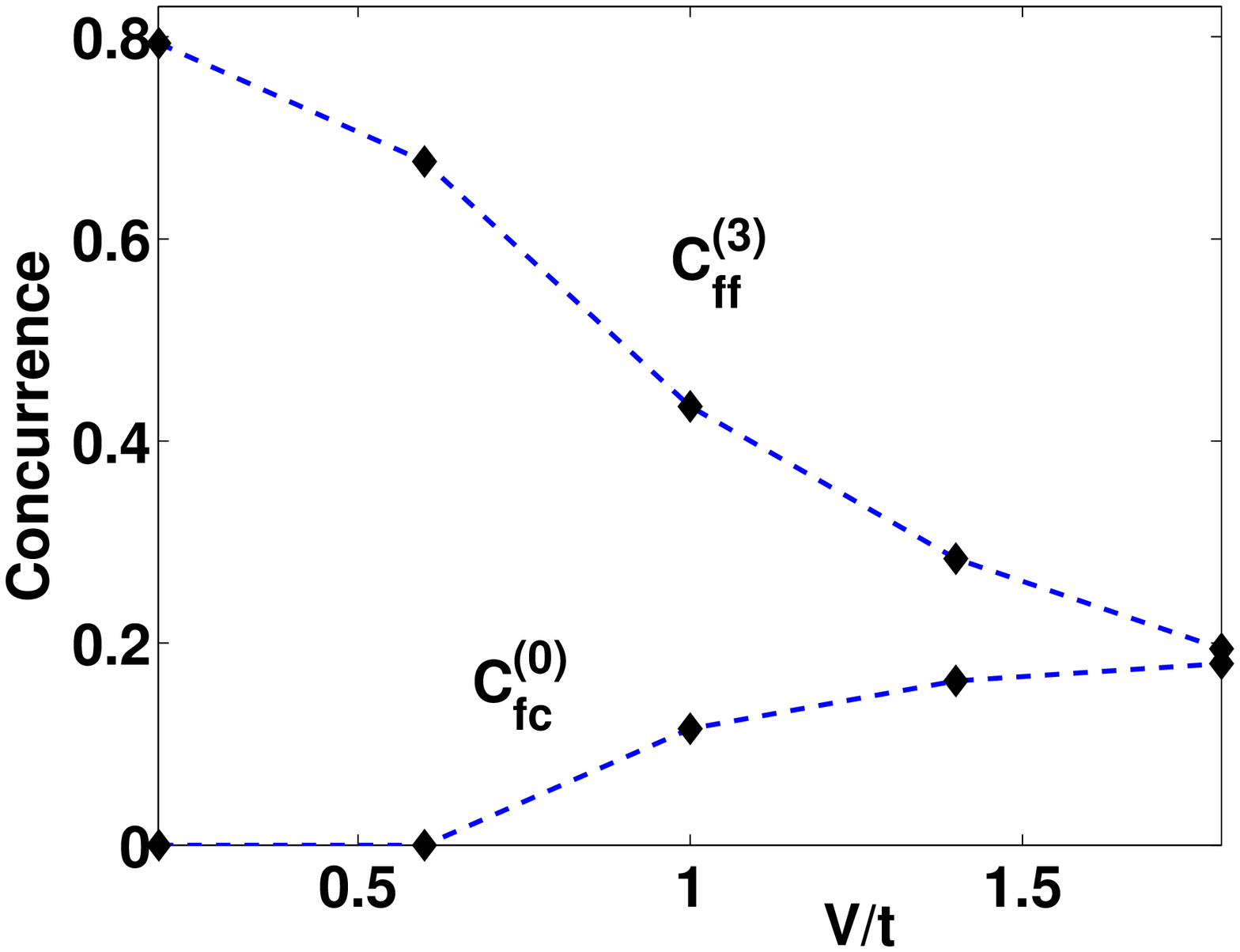,width=4cm}}
\caption{ Upper panels: The concurrence $C_{\text{ff}}^{(1)}$ (left)
and $C_{\text{fc}}^{(0)}$ (right) as a function of magnetic field $B$
and hybridization $V/t$. Lower left: The ff-concurrence
$C_{\text{ff}}^{(j)}$ for $j=1,2,3$ as a function of spin
$S^{\text{tot}}$. Lower right: The concurrence $C_{\text{ff}}^{(1)}$
and $C_{\text{fc}}^{(0)}$ as a function of hybridization $V/t$ for
$N_e=8$ electrons.  In all panels, $L=6$, $E_{\text{f}}=-U/2=-3t$ and
$T=0$.}
\label{fig3}
\vspace{-0.5cm}
\end{figure}

We plot (see Fig. \ref{fig3}) $C_{\text{ff}}^{(1)}$ and
$C_{\text{fc}}^{(0)}$ as a function of $B$ and $V/t$ for $L=6$. Since
$[\sum_{i\alpha}S_{i\alpha}^z,H]=0$ a finite B-field only shifts all
the manybody states in energy an amount proportional to their total
$S_z$ quantum number, eventually inducing a crossing of the $B=0$
levels and changing the ground state. This is clearly illustrated in
Fig. \ref{fig3}. On increasing $B$, the total ground state spin
$S^{\text{tot}}$ increments monotonically in steps of one, i.e. $B$
induces successive flipping of the spins. While for $V>t$ the
fc-entanglement is suppressed on increasing $B$, the ff-entanglement
survives for $V<t$ up to $S^{\text{tot}}=2$. Notably, for next and
second next n.n. sites, entanglement actually increases with
$S^{\text{tot}}$ and at $S^{\text{tot}}=2$ we have
$C_{\text{ff}}^{(j)}\approx 0.3$ for all $j=1,2,3$
(Fig. \ref{fig3}). The analysis of the exact wavefunction shows that
the ground state for $V\ll t$ is given by $|\Psi_{\text{g}}
\rangle=|\Psi_{\text{f}} \rangle\otimes|\Psi_{\text{c}} \rangle$, with
\begin{equation}
|\Psi_{\text{f}} \rangle=\sum_{i=1}^6 f^{\dagger}_{i\downarrow}\prod_{j\neq i}f_{j\uparrow}^{\dagger}|0\rangle
\label{Wstate}
\end{equation}
This is a W-state, just as was found in spin clusters. A magnetic
field thus flips the f-spins before flipping any c-spin, a consequence
of the much higher cost ($t$ compared to $0.01 t$) in energy to flip a
c-spin. In the W-state all particles are mutually entangled with the
same maximal, pairwise concurrence $C=1/3$ \cite{Coffman}, in good
agreement with the ED results in Fig. \ref{fig3}. The W-state is found
only for $L=6$; instead, for $L=3$ to $5$ there is negligible
entanglement away from $S^{\text{tot}}=0$, an indication of a strong
size dependence of this result.

{\it Away from half filling}.  For $N_{e}\neq 2L$, the PAM phase
diagram is rather complicated \cite{Sigrist}, in several parameter
regimes the ground state is magnetic, $S^{\text{tot}}>0$. To
investigate the effect of the electron concentration on the
entanglement, we considered fillings $2L-1\ge N_{e}\ge L+1$ for
$U=-2E_f=6t$, $B=0$ and $T=0$. We found that the overall behavior with
ff-entanglement at $V<t$ and fc-entanglement at $V>t$ prevails away
from half filling. For $V>t$, reducing $N_{e}$ monotonically
suppresses the fc-entanglement, due to an incomplete local Kondo
screening. Interestingly, for $V<t$ we find finite ff-entanglement
only for $L=4,6$ with $N_{e}=L+2$. Since $n_f\approx 1$, this
corresponds to two electrons in the conduction band. For this filling,
the KLM ground state exhibits ff-spin di- and trimerization for $L=4$
and $6$ respectively \cite{Affleck}. We indeed find that the many-body
ground state is dominated by the term $|\Psi \rangle=|\Psi
_{\text{f}}\rangle\otimes|\Psi_{\text{c}} \rangle$, with
$|\Psi_{\text{f}} \rangle$ a linear combination of $L$ dimerized
terms,
\begin{equation}
|\Psi_{\text{f}}
 \rangle=2^{-1}\left[|\bbb \downarrow\downarrow\uparrow\uparrow\rangle+|\bbb \uparrow\uparrow\downarrow\downarrow\rangle+|\bbb \downarrow\uparrow\uparrow\downarrow\rangle+|\bbb \downarrow\uparrow\uparrow\downarrow\rangle\right]
\label{dimstate}
\end{equation}
for $L=4$ and similar for $L=6$. This state has the property that only
sites opposite to each other in the ring show a finite entanglement,
$C_{\text{ff}}^{(L/2)}=4/L$, in reasonable agreement with the
numerical results (see Fig. \ref{fig3}).

In conclusion, we have studied two-body entanglement in Anderson
nanoclusters. We presented an entanglement phase diagram describing a
generic entanglement scenario for systems with Kondo-RKKY
competition. We also showed that Anderson nanoclusters exhibit
multiparticle entanglement depending on parameter regimes, electron
filling or applied magnetic fields. More generally, our results give
evidence that the interplay of charge and spin degrees of freedom must
be taken into account to assess the entanglement behavior of
nanoclusters with magnetic impurities.  We acknowledge useful
discussions with C-O. Almbladh.  This work was supported by the
Swedish VR (P.S.) and EU 6th framework Network of Excellence
NANOQUANTA (NMP4-CT-2004-500198) (C.V.).


\begin{thebibliography}{02}
\bibitem{NielsenChuang}
M. Nielsen and I. Chuang,  {\it Quantum Computation and Quantum Information}, (Cambridge University Press, Cambridge 2000).
\bibitem{Manoharan}
H.C. Manohoran, C.P. Lutz and D.M. Eigler, Nature {\bf 403}, 512 (2000).
\bibitem{Odom}
T.W. Odom {\it et al}., Science {\bf 290}, 1549 (2000).
\bibitem{Craig}
N.J. Craig {\it et al}.,  Science {\bf 301}, 565 (2004).
\bibitem{Doniach}
S. Doniach, Physica B+C {\bf 91}, 231 (1977). 
\bibitem{Sigrist}
H. Tsunetsugu, M. Sigrist and K. Ueda, Rev. Mod. Phys. {\bf 69} 809 (1997).
\bibitem{Kondoref}
G. R. Stewart, Rev. Mod. Phys. {\bf 73}, 797 (2001).
\bibitem{Kim}
S. Oh, J.Kim, Phys. Rev. B {\bf 73}, 052407 (2006);
E. S. S\o rensen \textit{et al}., cond-mat/0606705.
\bibitem{Cho}
S.Y. Cho, R.H. McKenzie, Phys. Rev. A {\bf 73}, 012109 (2006).
\bibitem{Saguia}
A. Saguia, M.S. Sarandy, Phys. Rev. A {\bf 67}, 012315 (2003). 
\bibitem{spinmod}
K. M. O'Connor, W. K. Wootters, Phys. Rev. A {\bf 63}, 052302 (2001);
X. Wang, H. Fu, and A.I. Solomon, J. Phys. A: Math. Gen. {\bf 34} 11307 (2001); 
I. Bose and A. Tribedi,  Phys. Rev. A  {\bf 72}, 022314 (2005). 
\bibitem{Arnesen}
M.C. Arnesen, S. Bose, and V. Vedral, Phys. Rev. Lett. {\bf 87}, 017901 (2001); G. L. Kamta and A.F. Starace, {\it ibid} {\bf 88}, 107901 (2002).
\bibitem{Gu}
S.-J. Gu, \textit{et al}., Phys. Rev. Lett  {\bf 93}, 086402 (2004).
\bibitem{Johannesson}
D. Larsson, H. Johannesson, Phys. Rev. Lett  {\bf 95}, 196406 (2005); 
Phys. Rev. A {\bf 73}, 042320 (2006).
\bibitem{Fazio}
A. Osterloh {\it et al}., Nature {\bf 416}, 608 (2002); 
T.J. Osborne and M.A. Nielsen, Phys. Rev. A {\bf 66}, 032110 (2002). 
\bibitem{Nozieres}
P. Nozieres, {\it Theory of Interacting Fermi Systems}, (Benjamin, N.Y. 1964).
\bibitem{Werner}
R. F. Werner, Phys. Rev. A {\bf 40}, 4277 (1989).
\bibitem{Wooters}
W. K. Wootters,  Phys. Rev. Lett  {\bf 80}, 2245 (1998).
\bibitem{conccom}
Since states with zero or two particles at an orbital do not contribute to the entanglement, we have $0 \leq C \leq 1$.
\bibitem{Wiseman}
H. M. Wiseman, J.A. Vaccaro, Phys. Rev. Lett  {\bf 91}, 097902 (2003).
\bibitem{verdozzidis}C. Verdozzi , Y. Luo, N. Kioussis, Phys.Rev. B {\bf 70}, 132404 (2004).
\bibitem{Verdozzi} 
Y. Luo, C. Verdozzi, N. Kioussis, Phys. Rev. B {\bf 71}, 033304 (2005). 
\bibitem{KLM2D}
I. Zerec, B. Schmidt and P. Thalmeier, Phys. Rev. B {\bf 73}, 245108 (2006). 
\bibitem{SchriefferWolf}
J.R. Schrieffer, P.A. Wolf, Phys. Rev. {\bf 149}, 491 (1966). 
\bibitem{tcom} 
Changing $t$, i.e. the conduction energy level spacing,
the entanglement phase diagram is
scaled in size.
\bibitem{Wstate}
W. D\"ur, G. Vidal, J.I. Cirac, Phys. Rev. A  {\bf 62}, 062314 (2000); X. Wang, {\it ibid} {\bf 64}, 012313 (2000).
\bibitem{Coffman}
V. Coffman, J. Kundu and W.K. Wootters, Phys. Rev. A {\bf 61}, 052306 (2000).
\bibitem{Affleck} 
J.C. Xavier {\it et al}, Phys. Rev. Lett. {\bf 90}, 247204 (2003),
Y. Chen and H. Chen, Phys. Rev. B {\bf 73}, 033402 (2006).


\end{thebibliography}
\end{document}